\documentclass[aps,pre,twocolumn,groupedaddress,superscriptaddress,showpacs]{revtex4-1}
\usepackage{epsfig,amssymb,amsmath,graphicx,subfigure,hyperref}
\usepackage{tabularx}
\usepackage{float}
\usepackage{setspace}
\usepackage{color}
\usepackage{times}
\usepackage{verbatim}

\usepackage{array} 
\newcommand{\PreserveBackslash}[1]{\let\temp=\\#1\let\\=\temp} \newcolumntype{C}[1]{>{\PreserveBackslash\centering}p{#1}} \newcolumntype{R}[1]{>{\PreserveBackslash\raggedleft}p{#1}} \newcolumntype{L}[1]{>{\PreserveBackslash\raggedright}p{#1}} 
\usepackage{textcomp}

\begin{document}

\title{The configurational entropy of colloidal particles in a confined space}

\author{Duanduan Wan}
\email[E-mail: ]{ddwan@whu.edu.cn}
\affiliation{School of Physics and Technology, Wuhan University, Wuhan 430072, China}

\date{\today}
\begin{abstract}
We calculate the configurational entropy of colloidal particles in a confined geometry interacting as hard disks using Monte Carlo integration method. In particular, we consider systems with three kinds of boundary conditions: hard, periodic and spherical. For small to moderate packing fraction $\phi$ values, we find the entropies per particle for systems with the periodic and spherical boundary conditions tend to reach a same value with the increase of the particle number $N$, while that for the system with the hard boundary conditions still has obvious differences compared to them within the studied $N$ range. Surprisingly, despite the small system sizes, the estimated entropies per particle at infinite system size from extrapolations in the periodic and spherical systems are in reasonable agreement with that calculated using thermodynamic integration method. Besides, as $N$ increases we find the pair correlation function begins to exhibit similar features as that of a large self-assembled system at the same packing fraction. Our findings may contribute to a better understanding of how the configurational entropy changes with the system size and the influence of boundary conditions, and provide insights relevant to engineering particles in confined spaces.  
\end{abstract}
\maketitle

\section{Introduction}
The structural and thermodynamical properties of colloidal particles in a confinement can be affected significantly by the shape and size of the confining boundary , especially when the number of particles is small, such as from a few to tens of particles (e.g., Refs.~\cite{Kravitz1969, Carlsson2012, Teich2016, Wan2018, Fonseca2019}). For example, two-dimensional packings of hard spheres under circular confinement results in optimal circle packings forming doublets, triangles, squares, pentagons, and hexagons \cite{Kravitz1969}. Another example is with five disks in the unit square, the topology changes at least 20 times as the disk radius varies \cite{Carlsson2012}. In fact, in a three-dimensional (3D) $N$-particle system with free boundaries, the fraction of all particles that is at the surface is proportional to $N^{-1/3}$ \cite{Frenkel1996}. Thus in the case of studying bulk phases, in order to reduce surface effects and simulate systems in the thermodynamic limit, most simulations probe a system of a larger size, e.g.~a few hundred to a few thousand particles, and use periodic boundary conditions that mimic the presence of an infinite bulk surrounding the studied system (e.g., Refs.~\cite{Torquato2009, Haji-Akbari2009, Damasceno2012, Avendano2012, deGraaf2012, Ni2012, Gantapara2013, Klotsa2018, Wan2019, Wan2021}). 

Here we attempt to explore the influence of boundary conditions as a function of the system size more quantitatively. For simplicity, we study 2D hard disks which serve as a model system of spherical colloidal particles constrained to a plane. 
We consider hard and periodic boundary conditions respectively. The hard boundary conditions is adapted as a square box with hard walls. The periodic boundary condition is implemented as a square box with periodic boundary conditions,  and once a part of a particle is outside the 2D periodic box, the outside part will re-appear through the opposite side of the box. In this way the system has the topology of a torus. Besides, we also consider a spherical topology to realize the periodic boundary condition, as a 2D system may be embedded in the surface of a sphere without introducing any physical boundaries \cite{Hansen1979}. Thus the comparison between the periodic and spherical cases may allow us to explore the effects of the underlying topology.

In this work, we calculate the configurational entropy (entropy associated with particle positions) of the system with the three kinds of boundary conditions mentioned above using Monte Carlo (MC) integration method. We find the entropies per particle for systems with the periodic and spherical boundary conditions already tend to reach a same value despite the small system size (up to the number of particles $N \lessapprox 100$ in the lowest packing fraction case), while that for the system with the hard boundary conditions has obvious differences compared to them. Surprisingly, despite the small system sizes, the estimated entropies per particle at infinite system size from extrapolations in the periodic and spherical systems are in reasonable agreement with that calculated using thermodynamic integration method. Further, as $N$ increases we find the pair correlation function begins to exhibit similar features as that of a large self-assembled system at the same packing fraction. We discuss the influence of boundary conditions and related systems of engineering particles in a confined space.

\section{Methods}

\begin{figure}[h]
\centering 
\includegraphics[width=3.5in]{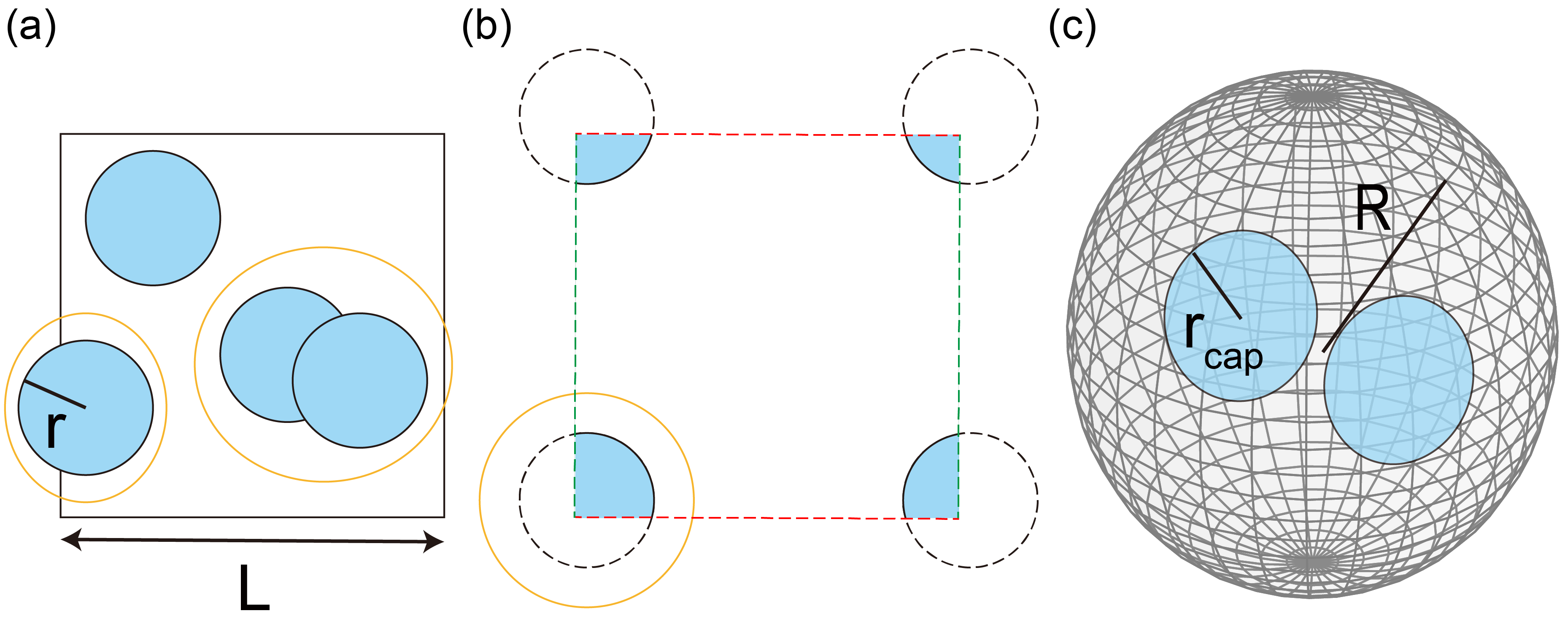} 
\caption{(Color online) Three boundary conditions considered in this work: hard, periodic and spherical boundary conditions. Particles (light blue) are of unit area. (a) A square box with hard boundary conditions.  Two failure situations (orange circles): particles partly outside the boundary and overlapping with a previous particle. (b) A square box with periodic boundary conditions. The red (green) dashed lines indicate two opposite sides of the periodic boundaries. Parts of a particle (orange circle) which lie outside the boundary sides re-appear through the opposite sides. (c) Periodic boundary conditions with a spherical topology. Particles are considered as spherical caps.}
\label{three_methods}
\end{figure}

We adapt the MC integration method in Ref.~\cite{Wan2018} to calculate the configurational entropy. We briefly describe the idea here. To perform the MC integration with the hard boundary conditions, we first placed a particle into a square box with its position randomly selected inside the box. If the particle lies completely inside the box, we randomly placed a second particle. If the second particle is also inside the box boundary and does not overlap with the first one, we placed a third particle. This procedure continued until one of two failure situations shown in Fig.~\ref{three_methods}(a) occurred. At a given packing fraction $\phi$, we calculate the entropy as a function of the particle number $N$. We fixed the area of the hard disks to be 1, i.e. the radii of the hard disks are $r=\sqrt{1/\pi}$. Thus the side length of the square box is $L=\sqrt{N/\phi}$. For each particle number $N$ we ran $n_{tot}$ trials and counted the number of successful trials $n_{s}$. The configurational free volume $V(N)$ of $N$ particles then is:
\begin{equation}
V(N)=\frac{n_{s}(N)}{n_{tot}}\times A_{box}^{N},
\label{Volume}
\end{equation}
where $A_{box}$ is the area of the box. For the system with the periodic boundary conditions, similarly, we placed particles randomly inside the box. The difference compared to the hard boundary condition case is if a part of a disk lies outside a side of the box, we consider this part re-appears through the opposite side (see an example in Fig.~\ref{three_methods}(b)). In this way particles always lie completely inside the box. Therefore, there is only one failure situation: the overlap between particles. This is decided by if there is overlap between any part of two disks. For the system with the spherical boundary conditions, particles were randomly placed on a sphere and were considered as spherical caps of unit area (see Fig.~\ref{three_methods}(c)). To generate uniformly distributed particles on a sphere, we obtained the polar angle of every particle using $\theta= \arccos (1-2u)$, where $u$ is a random number from the uniform distribution $U[0,1]$, and the azimuthal angle $\psi$ is a random number from $U[0,2\pi]$.
For a given packing fraction $\phi$ value, the radius of the underlying sphere is $R=\sqrt{N / 4 \pi \phi }$ and the radius of the spherical cap is $ r_{cap}=R  \arccos (1-1/2 \pi R^{2})$. The overlap between two spherical caps is determined by whether the length of the geodesic between the centers of the spherical caps is shorter than $2r_{cap}$. The $A_{box}$ in Eq.~(\ref{Volume}) is replaced by $A_{sphere}$ for the spherical boundary condition case. We take $n_{tot}=10^{10}$ for all calculations. The entropy of the particles relative to that of an ideal gas is:
\begin{equation}
S(N)= k_{B} \mbox{ln}\frac{V(N)}{V_{0}(N)},
\label{S}
\end{equation}
where $k_{B}$ is Boltzmann's constant (we set $k_{B}=1$), $V_{0}(N)$ is for normalization and $V_{0}(N)= A_{box}^{N}$ (or $A_{sphere}^{N}$) is the free volume of $N$ ideal gas particles in the box (or on the sphere). This normalization makes $S(N)$ independent of the length scale of the system. As $S(N)=0$ for an ideal gas, the entropies we calculate here are negative. It follows that the entropy per particle is $s(N)= S(N)/N$.

\section{Results and discussion}

\begin{figure}
\centering 
\includegraphics[width=3in]{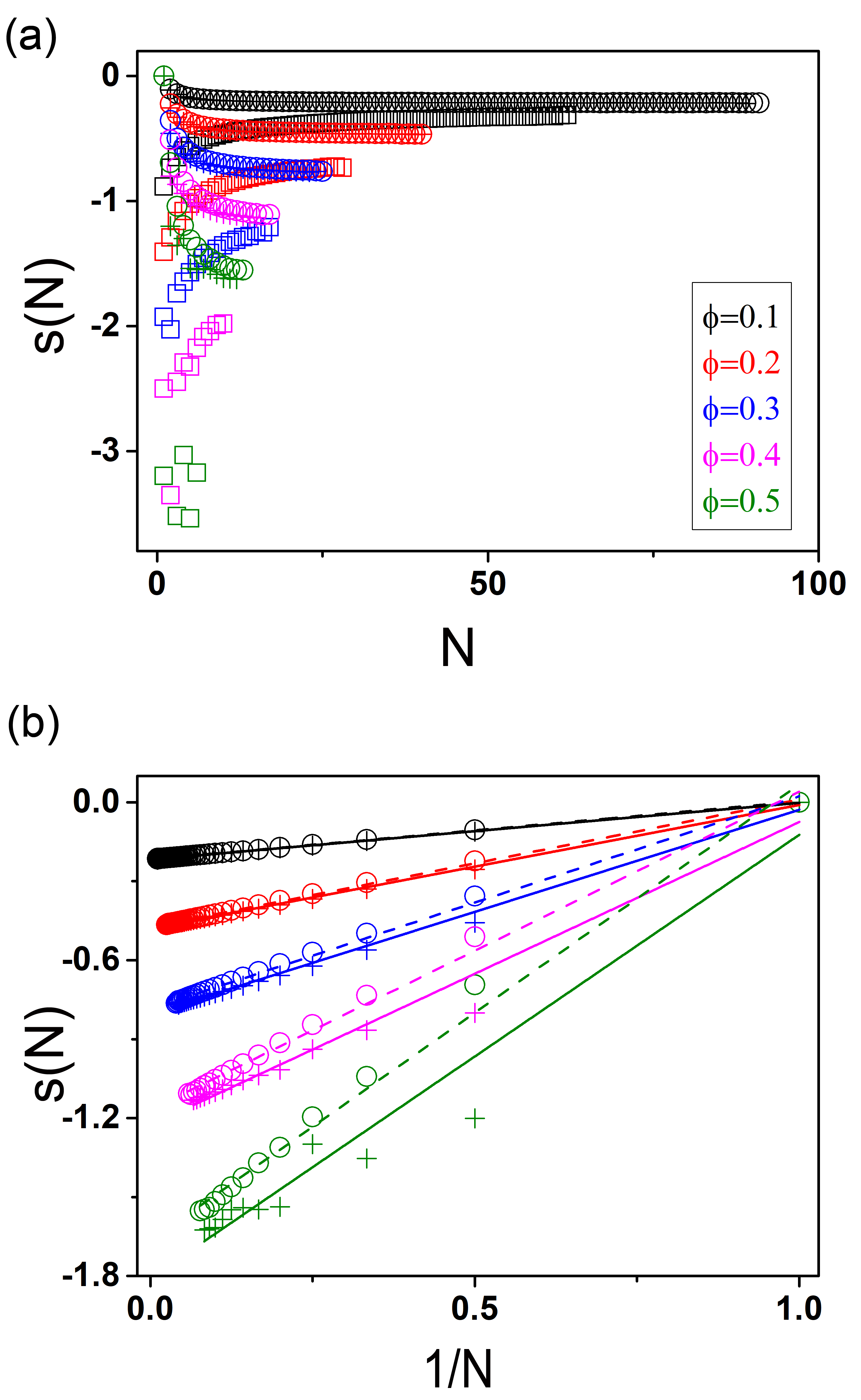} 
\caption{(Color online) (a) Entropy per particle $s(N)$ as a function of particle number $N$ at various $\phi$ values. Square symbols, plus signs and circles represent systems with hard, periodic and spherical boundary conditions, respectively. (b) The same plots of $s(N)$ for systems with the periodic and the spherical boundary conditions, but as a function of $1/N$. Lines are linear fits of the symbols, with the straight lines for the periodic system and dashed lines for the spherical system. Fitting coefficients are shown in Table~\ref{coefficients}.}
\label{entropy_collection}
\end{figure}

\begin{figure*}
\centering 
\includegraphics[width=7in]{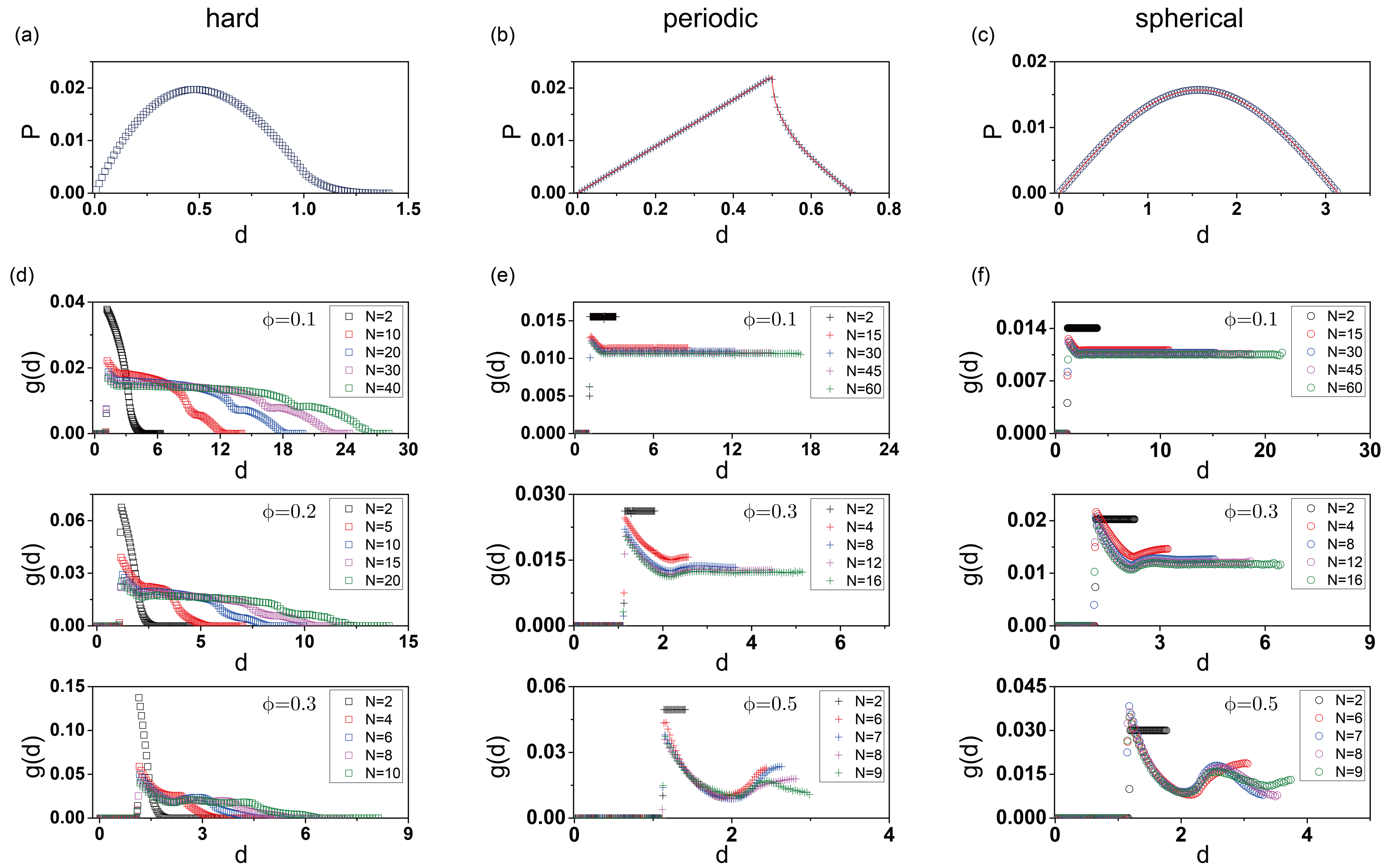} 
\caption{(Color online) (a-c) Probability density of finding a particle a distance $d$ away from another particle for an ideal gas, with the hard, periodic and spherical boundary conditions and length scales $L=1$, $L=1$ and $R=1$ in the three cases, respectively. Red lines in (b) and (c) are theoretical expectations using Eq.~(\ref{eq_periodic}) and Eq.~(\ref{eq_spherical}). (d-f) Pair correlation function (defined as the probability of finding a particle at a distance of $d$ away from another particle, relative to that for an ideal gas) for the three kinds of boundary conditions at some selected $\phi$ and $N$ values.}
\label{pair_correlation}
\end{figure*}

\begin{figure}
\centering 
\includegraphics[width=3in]{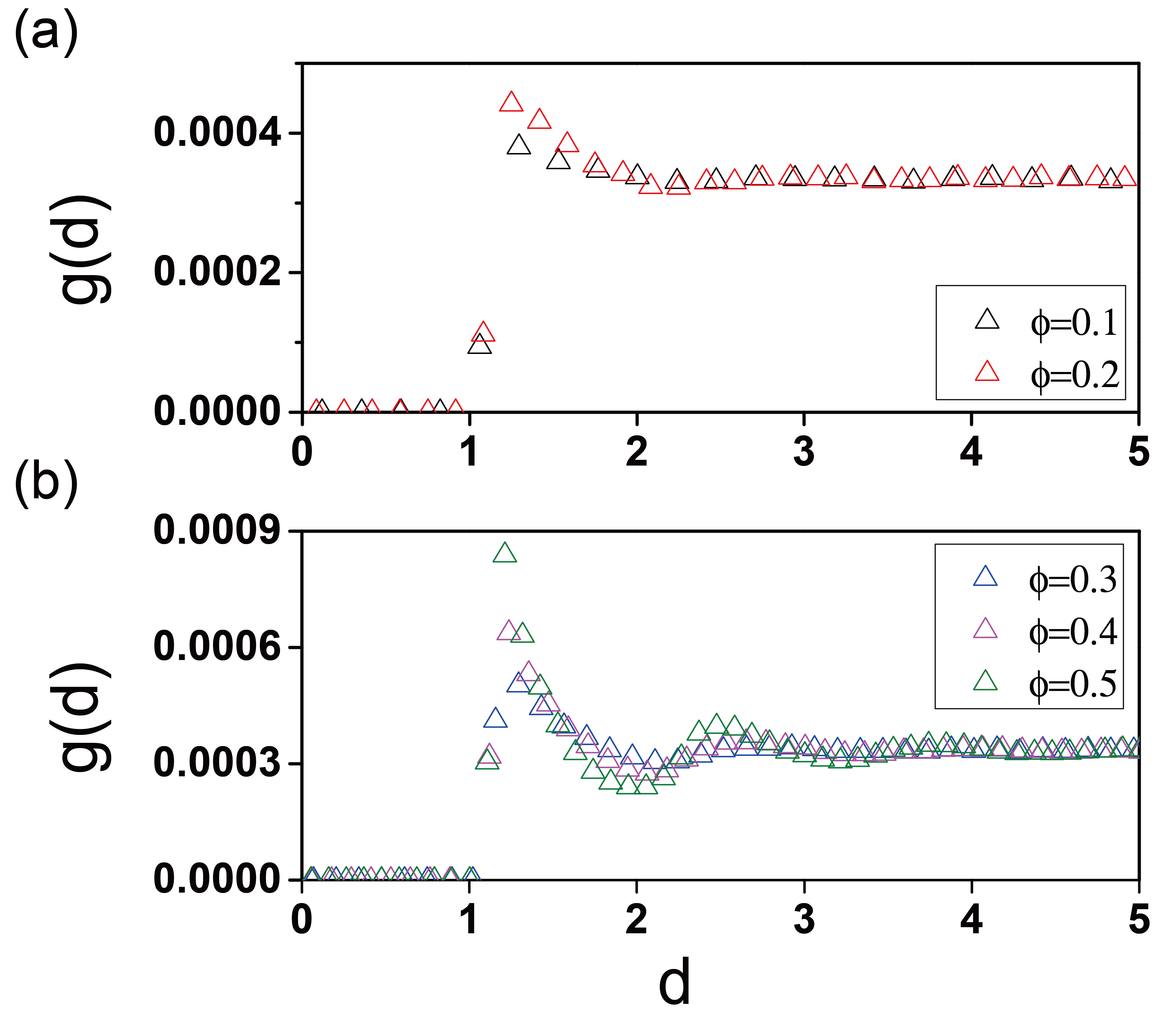} 
\caption{(Color online) Pair correlation function for a self-assembled system with $N=10^5$ particles, with a short distance of $d \in [0, 5]$ shown.}
\label{sa}
\end{figure}

We plot $s(N)$ as a function of $N$ at fixed packing fraction $\phi$ values in Fig.~\ref{entropy_collection}. Trials that achieved larger $N$ or $\phi$ values were rare and thus the errors are relatively large (see the SI for more information). The plot shows several essential features: $s(N)$ of the periodic and the spherical systems in general show a descending trend as $N$ increases, while $s(N)$ with the hard boundary conditions shows a rising trend. For a same $\phi$ and $N$ values, as expected, $s(N)$ of the periodic and the spherical systems are always larger than that with the hard boundary conditions and the differences decrease as $N$ increases. Besides, for $\phi \le 0.3$, $s(N)$ of the periodic and the spherical systems already agree well at the $N$ value of a few tens. 

If plotting $s(N)$ of the periodic and the spherical systems as a function of $1/N$ instead (Fig.~\ref{entropy_collection}(b)), it is more obvious that $s(N)$ of the spherical system is slightly larger than that of the periodic system. This difference can be partly due to the different underlying topologies of the two systems. Besides, that the difference decreases as $N$ increases suggests the influence of the underlying topology decreases as the system size increases. Particles arranged on the surface of a sphere are known to exhibit a richness of defect structures in a variety of systems (e.g.~\cite{Thomson1904, Caspar1962, Bowick2002, Lidmar2003, Bowick2006, Wales2006, Brojan2015, Wan2015, Mehta1016, Yao2017, Yao2019}), which is a result of both the inter-particle interaction and the spherical topology. Reference \cite{Miller2011} finds the phase diagram and structures of purely repulsive soft colloidal particles constrained to a two-dimensional plane and to the surface of a spherical shell using numerical simulations. In practice, experiments have realized colloidal particles confined to a spherical surface by beads self-assembled on water droplets in oil \cite{Bausch2003}. As $s(N)$ of the periodic and the spherical systems show a linear trend, we thus attempted to do a linear extrapolation of the data. The fitting coefficients are shown in Table~\ref{coefficients}. The intercepts thus correspond to estimated $s(N)$ values at an infinite system size (i.e., $N \rightarrow \infty$). We have also attempted to estimate $s(N)$ using thermodynamic integration method. The free energy of the hard disk gas can be computed from that of the ideal gas using the thermodynamic relation in Ref.~\cite{Frenkel1996}:
\begin{equation}
\frac{F(\rho)}{Nk_{B}T}= \frac{F^{id}(\rho)}{Nk_{B}T} + \frac{1}{k_{B}T} \int^{\rho}_{0} d\rho' (\frac{P(\rho')-\rho' k_{B}T}{\rho'^{2}}),
\label{free_energy}
\end{equation}
where $\rho=N/V$ is the particle density, $F^{id}(\rho)$ is the free energy for the ideal gas and $P(\rho')$ is the pressure of the hard disk system. Taking the virial coefficients of hard disks to the tenth to calculate the pressure \cite{Clisby2006}, we obtain the free energy difference per particle $f(\phi)/T=(F(\phi)-F^{id}(\phi))/NT$ (with particle area 1 and $k_{B}=1$ as above). As both the ideal gas and hard disk systems have the same kinetic energy at a given temperature and have zero potential energies, $f(\phi)/T$ is expected to equal to $-s(\phi)$. We put $-f(\phi)/T$ values in Table~\ref{coefficients} as well. As can be seen from the table, the intercepts in both the periodic and the spherical systems show reasonable agreement with the $-f/T$ values.

\begin{table}
\caption{The estimated entropy per particle $s(N)$ from MC integration method and from thermodynamic integration method.}
\begin{center}
\begin{tabular}{c c c c c} 
 \hline \hline
$\phi$ & & periodic system & spherical system &  $-f/T$ \\ [1ex] 
 \hline
 0.1 &  intercept   &  -0.21751  & -0.21683   & -0.21721\\
     &  slope       &   0.21527  &  0.21936   & \\
 0.2 &  intercept   &  -0.47906  &  -0.4739  & -0.47655\\
     &  slope       &   0.46872  &  0.48398   & \\
 0.3 &  intercept   &  -0.80434  &  -0.78474  & -0.79423\\
     &  slope       &   0.77558  &  0.80764   & \\
 0.4 &  intercept   &  -1.22786  &  -1.16871  & -1.19629\\
     &  slope       &   1.15242  &  1.20905   & \\
 0.5 &  intercept   &  -1.80831  & -1.66654   & -1.72674\\
     &  slope       &   1.6843   &  1.73392   &\\
\hline
\end{tabular}
\end{center}
\label{coefficients}
\end{table}

We also take a look at the particle probability distribution (local density). In the system with the hard boundary conditions, for $N = 1$ the density is uniform except at the boundary where it decreases abruptly to zero; for $N>1$, as $\phi$ increases, the density begins to increase in the four corners of the square cavity compared to elsewhere in the cavity. This tendency is the same as that in Ref.~\cite{Wan2018}. In the periodic and the spherical systems, the density is always uniform. 

We continue to explore the pair correlation function. In simulations we calculated the distance between all particle pairs and binned them in 100 equally spaced bins in the range of zero to the maximal length scale of the particle distance, i.e., $\sqrt{2}L$, $\sqrt{2}L/2$ and $\pi R$ ($L$ is the side length of the square boxes and $R$ the radius of the sphere) in the three cases, respectively. For the ideal gas system, Fig.~\ref{pair_correlation}(a-c) show the probability density of finding a particle a distance $d$ away from another particle in the three cases. For the ideal gas with the hard boundary conditions, an analytical expression for the probability density distribution is unknown. For the ideal gas with periodic boundary conditions, it reads
\begin{equation}
P(d)=\left\{
\begin{aligned}
&2\pi d, & 0 < d \le L/2 \\
&[2\pi - 8 \arccos (L/2d) ]d, &  L/2 < d \le \sqrt{2}L/2 \\ 
\end{aligned} 
\right.
\label{eq_periodic}
\end{equation}
where $L$ is the side length of the periodic box. 
For the spherical case, 
\begin{equation}
P(d)= 2 \pi R \sin(d/R),  \, 0 < d < \pi R, 
\label{eq_spherical}
\end{equation}
where $R$ is the radius of the sphere. In the latter two cases, the analytical expressions fit the numerical results well (Fig.~\ref{pair_correlation}(b-c)). 
We then consider the pair correlation function $g(d)$ of the hard disk system. The statistics was normalized with respect to the ideal gas, where we used the numerical result for the hard boundary condition case and analytical expressions for the periodic and spherical cases for normalization. Figure \ref{pair_correlation}(d-f) show $g(d)$ for the hard particle systems at some selected $\phi$ and $N$ values. As expected, $g(d)$ has a discontinuity at the distance of particle diameter. In the case of the hard boundary conditions, $g(d)$ decays to zero when the distance approaches the largest length scale (i.e.~$\sqrt{2}L$) of the system. In the cases of the periodic and spherical boundary conditions, at a relatively low $\phi$ value (e.g.,~$\phi \le 0.3$), $g(d)$ has a sharp first peak and then tends to be constant as the distance increases; at a larger $\phi$ value (e.g.,~$\phi = 0.4$ and $\phi=0.5$), $g(d)$ begins to show multiple peaks. The behavior of $g(d)$ in the systems with the periodic and the spherical boundary conditions is similar to that in a self-assembled hard disk system for short distances. For example, Fig.~\ref{sa} shows $g(d)$ of the self-assembled structures of $N=10^5$ hard disks at same $\phi$ values. The self-assembled structures were generated using the method in Ref.~\cite{Michael2016} and periodic boundary conditions were considered. To plot $g(d)$ for this larger system, we used 3000 equally spaced bins in the range of $[0, \sqrt{2}L/2]$ and Fig.~\ref{sa} shows a short-distance part. 

\section{Conclusions}
We used MC integration method to calculate the configurational entropy of colloidal particles in a confined space, with hard, periodic and spherical boundary conditions, respectively. For small to moderate packing fraction $\phi$ values, we find the entropies per particle for systems with the periodic and spherical boundary conditions tend to reach a same value with the increase of the particle number $N$, while that for the system with the hard boundary conditions still has obvious differences compared to them within the studied $N$ range. Besides, the estimated entropies per particle at infinite system size from extrapolations in the periodic and spherical systems are in reasonable agreement with that calculated using thermodynamic integration method. Further, as $N$ increases we find the pair correlation function begins to exhibit similar features as that of a large self-assembled system at the same packing fraction. Our findings may shed light on how the entropy per particle and the influence of boundary conditions change with the increase of the system size, and provide insights relevant to engineering particles in confined spaces, such as engineering colloidal particles in a confined flat plane or on a spherical surface.  

\acknowledgments  
The author thanks Yuheng Yang for checking the free energy difference independently. This work was supported by the National Natural Science Foundation of China (Grant No.~11904265), the Hubei Provincial Natural Science Foundation (Grant No.~2020CFB670) and the Fundamental Research Funds for the Central Universities (Grant No.~2042020kf0033). 

\bibliography{entropy}

\newpage
\clearpage
\onecolumngrid
\begin{center}
\textbf{\large Supplementary Material}
\end{center}
\setcounter{equation}{0}
\setcounter{figure}{0}
\setcounter{table}{0}
\setcounter{page}{1}
\makeatletter
\renewcommand{\theequation}{S\arabic{equation}}
\renewcommand{\thefigure}{S\arabic{figure}}

\begin{figure}[h]
\centering 
\includegraphics[width=3in]{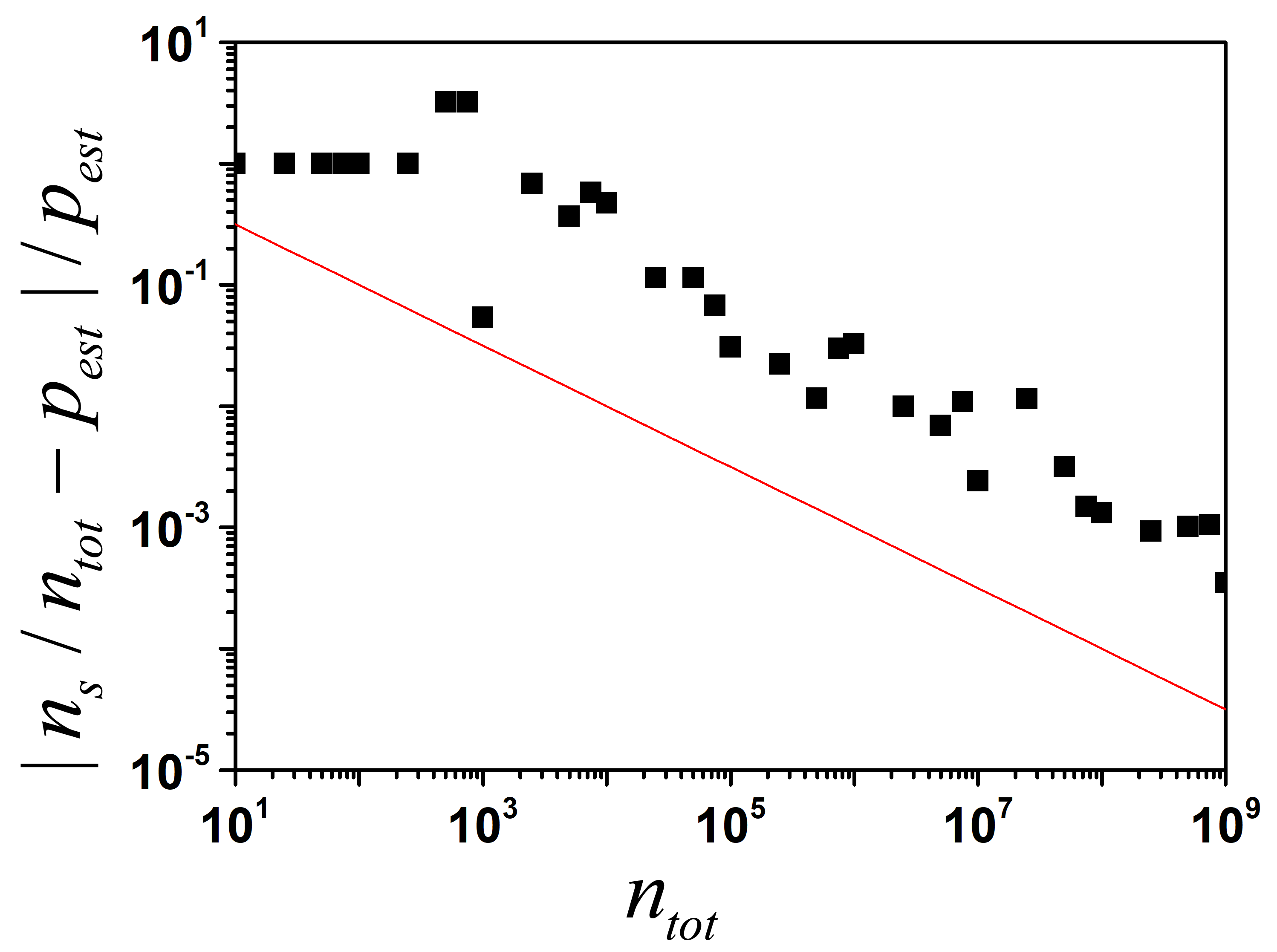}
\caption{(Color online) Estimated errors as a function of $n_{tot}$ in the periodic system with $\phi=0.1$ and $N=33$. The estimated probability $p_{est}$ is $n_{s}/n_{tot}$ value at $n_{tot}=10^{10}$. The red line plots $1/\sqrt{n_{tot}}$.}
\label{converge}
\end{figure}

We show a typical example of how estimated errors change with $n_{tot}$ in Fig.~\ref{converge}. We use the probability with $n_{tot}=10^{10}$ as the estimated probability $p_{est}$ and plot errors compared to it. The errors reduce as $1/\sqrt{n_{tot}}$, as expected for MC integration method. In the main text, for a given $\phi$ value, at each $N$ value we compare $s(N)$ values from two independent runs with $n_{tot}=10^9$ and $n_{tot}=10^{10}$. We truncate at the $N$ value where the two $s(N)$ values has an error larger than $3\%$.    

\end{document}